\documentclass[]{emulateapj}
\usepackage{graphicx}
\usepackage{natbib}
\usepackage{amssymb,amsmath}

\newcommand{\wise}{\textit{WISE}}

\newcommand{\lbol}{$L_{\textrm{bol}}$}
\newcommand{\OIII}{[O\,{\sc iii}]}

\begin{document}

\title{The \OIII\ profiles of infrared-selected active galactic nuclei: More powerful outflows in the obscured population}
\author{M.A. DiPompeo\altaffilmark{1}, R.C. Hickox\altaffilmark{1}, C.M. Carroll\altaffilmark{1}, J.C. Runnoe\altaffilmark{2}, J.R. Mullaney\altaffilmark{3}, T.C. Fischer\altaffilmark{4}}
\altaffiltext{1}{Department of Physics and Astronomy, Dartmouth College, 6127 Wilder Laboratory, Hanover, NH 03755, USA}
\altaffiltext{2}{Department of Astronomy, University of Michigan, 1085 S. University, Ann Arbor, MI 48109, USA}
\altaffiltext{3}{Department of Physics and Astronomy, The University of Sheffield, Hounsfield Road, Sheffield S3 7RH, UK}
\altaffiltext{4}{Astrophysics Science Division, Goddard Space Flight Center, Code 665, Greenbelt, MD 20771, USA}

\keywords{galaxies: active; galaxies: evolution; (galaxies:) quasars: general}

\begin{abstract}
We explore the kinematics of ionized gas via the [O III] $\lambda$5007 emission lines in active galactic nuclei (AGN) selected on the basis of their mid-infrared (IR) emission, and split into obscured and unobscured populations based on their optical-IR colors. After correcting for differences in redshift distributions, we provide composite spectra of spectroscopically and photometrically defined obscured/Type 2 and unobscured/Type 1 AGN from 3500 to 7000 \AA. The IR-selected obscured sources contain a mixture of narrow-lined Type 2 AGN and intermediate sources that have broad H$\alpha$ emission and significantly narrower H$\beta$.  Using both \OIII\ luminosities and AGN luminosities derived from optical-IR spectral energy distribution fitting, we find evidence for enhanced large-scale obscuration in the obscured sources. In matched bins of luminosity we find that the obscured population typically has broader, more blueshifted \OIII\ emission than in the unobscured sample, suggestive of more powerful AGN-driven outflows. This trend is not seen in spectroscopically classified samples, and is unlikely to be entirely explained by orientation effects. In addition, outflow velocities increase from small to moderate AGN $E(B-V)$ values, before flattening out (as traced by FWHM) and even decreasing (as traced by blueshift). While difficult to fully interpret in a single physical model, due to both the averaging over populations and the spatially-averaged spectra, these results agree with previous findings that simple geometric unification models are insufficient for the IR-selected AGN population, and may fit into an evolutionary model for obscured and unobscured AGN.
\end{abstract}

\section{INTRODUCTION} \label{sec:intro}
Observational evidence linking the fundamental properties of growing supermassive black holes (active galactic nuclei, AGN) and the galaxies that host them is firmly established \citep[e.g.][and references therein]{2013ARA&A..51..511K}.  This includes correlations between black hole mass and galaxy properties such as stellar velocity dispersion, bulge mass, and bulge luminosity \citep[][]{2000ApJ...539L...9F, 2000ApJ...539L..13G, 2003ApJ...589L..21M, 2004ApJ...604L..89H, 2007ApJ...663...61L, 2009ApJ...698..198G}.  From a theoretical standpoint, these links may be at least partially due to AGN ``feedback,'' in which energy is transported from the nucleus into the host, halting galaxy growth and quenching star formation (e.g.\ \citealt{2005Natur.433..604D, 2009MNRAS.398...53B, 2015MNRAS.448.1835T}; but see also \citealt{2013MNRAS.433.3079Z, 2013ApJ...774...66Z}).

Feedback operates broadly in two general ways: by injecting energy into the larger environment and halting the infall of gas onto galaxies, and by clearing gas already within the galaxy and quenching (or inducing) star formation.  The latter mechanism in particular however has mixed observational support \citep[][]{2009ApJ...706..525M, CanoDiaz:2012p2839, 2012ARA&A..50..455F, 2015ApJ...800...19R, 2015ApJ...799...82C, 2016A&A...585A.148B, 2016A&A...591A..28C, 2017ApJ...837...91B}. One potential mode of feedback is via outflows driven by the central AGN. Observations of large-scale galactic outflows driven by AGN have been identified at a range of redshifts in both molecular gas \citep[e.g.][]{Feruglio:2010p1255, 2014A&A...562A..21C, 2017arXiv170903510V, 2017arXiv170600443V} as well as low density ionized gas in the narrow-line region (NLR), which is ionized by the AGN and extends over kiloparsec scales \citep[e.g.][]{1989ApJ...345..730P, 2013ApJ...774..145H, 2014MNRAS.441.3306H}.

The bright \OIII\ $\lambda$5007 emission line is a popular diagnostic of the gas kinematics in AGN hosts, and broad, asymmetric components beyond that expected from galaxy rotation have been known for some time \citep[e.g.][]{1981ApJ...247..403H, 1982ApJ...256..427F, 1984ApJ...281..525H}. Both detailed analyses of spatially-resolved emission in small samples as well as statistical studies with larger spectroscopic datasets across a range of parameter space have been used to explore outflows in AGN, and all generally agree that outflows are ubiquitous.  The outflows traced by \OIII\ generally correlate with AGN luminosity in the optical, IR, and X-rays \citep[][]{2011ApJ...737...71Z, 2012MNRAS.426.1073H, 2014MNRAS.442..784Z, 2016MNRAS.459.3144Z, 2017ApJ...834...30F, 2017A&A...603A..99P, 2017ApJ...835..222S}, suggestive that radiation pressure is driving the winds.  In some cases however, outflows may be driven by radio jets \citep[e.g.][]{2006ApJ...650..693N, 2010ApJ...716..131R, 2013MNRAS.433..622M}, or radio emission is produced by shocks from a radiatively driven wind \citep[][]{2014MNRAS.442..784Z}.  Outflow velocities may also correlate with other AGN properties such as black hole mass \citep[][]{2017arXiv170805139R} and accretion rate \citep[][]{2005ApJ...627..721G, 2006MNRAS.372..876B, 2016ApJ...817..108W}, though in some cases evidence for this is weak at best \citep[][]{2011ApJ...737...71Z, 2014RAA....14..913P}. Several studies have found that winds traced by \OIII\ are energetic enough to provide significant feedback on the host galaxy gas supply on large scales \citep[][]{2010MNRAS.402.2211A, 2011ApJ...732....9G, 2012MNRAS.426.1073H, 2016MNRAS.459.3144Z}, though low luminosity systems may not be able to drive winds on sufficiently large scales \citep[e.g.][]{2017ApJ...834...30F}. In some cases \citep[e.g.][]{2016MNRAS.459.3144Z, 2017A&A...603A..99P}, outflows are suggested to be linked to the ``blow-out'' phase critical to many evolutionary models for AGN \citep[e.g.][]{2008ApJS..175..356H}.

A barrier in the study of the AGN-galaxy connection has been that a large fraction of luminous AGN are obscured by gas and dust, and missed by traditional (e.g.\ optical) photometric selection techniques that identify AGN in large numbers \citep[e.g.][]{2015ApJ...804...27A}. There are several ways of defining ``obscured'' and ``unobscured'' AGN, one of the most standard (and that used by the majority of the studies discussed above) being optical spectral classification.  In this case, unobscured or ``Type 1'' AGN generally have blue power-law continua and both broad ($\gtrsim 1000$ km/s) and narrow emission lines, while obscured or ``Type 2'' AGN typically have galaxy-like continua and only narrow emission lines. The obscuration in this classification is often attributed to dust that blocks a direct view of the accretion disk and broad line region, but leaves the larger-scale NLR visible \citep[e.g.][]{1993ARA&A..31..473A, Urry:1995p507}. However, there may also be an evolutionary component to this dichotomy \citep[e.g.][]{1984MNRAS.211P..33P}, illuminated more recently by so-called ``changing-look'' quasars \citep[e.g.][]{2015ApJ...800..144L, 2016MNRAS.455.1691R}. Because AGN samples selected for spectroscopic follow-up from optical photometric properties tend to rely on the strong blue colors of AGN, there are inherent biases in studies that rely on Type 1 and Type 2 samples from such surveys.

The hot dust in the nuclear regions that likely contributes to obscuration also allows efficient identification of AGN in the infrared (IR), methods of which were first developed with \textit{Spitzer} data \citep[e.g.][]{2004ApJS..154..166L, Stern:2005p2563}. These were later expanded to the entire sky with the \textit{Wide-field Infrared Survey Explorer} \citep[\wise;][]{2010AJ....140.1868W, 2012ApJ...753...30S, 2013ApJ...772...26A, 2013MNRAS.434..941M}. \citet{2007ApJ...671.1365H} found that the optical-IR colors of IR-selected AGN were bimodal, indicative of the presence of distinct populations (see Section~\ref{sec:wise_agn}).  Indeed, spectral energy distribution (SED) modeling shows all are well-fit by a superposition of galaxy and AGN templates, with the red and blue sources having similar AGN luminosities but significant extinction of the nuclear optical light accounting for the red colors in some of the sample \citep{2007ApJ...671.1365H, 2017ApJ...849...53H}. In addition, optical spectra of optically faint, IR-selected AGN are generally consistent with Type 2 AGN \citep{2013ApJS..208...24L,  2014ApJ...795..124H, 2017arXiv171101269H}, and the X-ray properties of such sources also agree with this interpretation \citep{2007ApJ...671.1365H}. Therefore, the red optical-IR colors in these samples are generally not simply due to an IR excess compared to the optical --- rather, optical-IR colors can be used as another method to define obscured and unobscured AGN.  

Throughout the remainder of this work, we will use the terms ``Type 1'' and ``Type 2'' to refer to sources classified via optical spectroscopy, and ``obscured'' and ``unobscured'' to refer to those categorized by optical-IR photometric colors.

The expansion of IR AGN selection across the whole sky with \wise\ provides a unique opportunity to study obscured and unobscured AGN selected in a similar way, with large numbers of both. Intriguingly, \wise-selected obscured AGN are found to have higher star formation rates \citep{2015ApJ...802...50C} and reside in more massive dark matter halos \citep{2011ApJ...731..117H, 2014MNRAS.442.3443D, 2015MNRAS.446.3492D, 2016MNRAS.456..924D, 2017MNRAS.469.4630D}, which may be broadly explained in an evolutionary framework \citep{2017MNRAS.464.3526D} consistent with some theoretical models \citep[e.g.][]{Hopkins:2006p654, 2008ApJS..175..356H}. This suggests that \wise-selected obscured AGN may represent a distinct phase in the lifetimes of quasars (or a mix of more traditionally unified objects, e.g.\ \citealt{1993ARA&A..31..473A, 2015ARA&A..53..365N} and others, e.g.\ \citealt{2016MNRAS.460..175D}).

If IR-selected obscured AGN probe a different evolutionary phase in which the AGN is transitioning from a obscured source to an unobscured AGN by blowing out dusty material \citep[e.g.][]{2012NewAR..56...93A}, we might expect to see differences in their gas kinematics on large scales as winds launched at different times in the AGN life cycle may have very different properties \citep{2017arXiv170403712I}. In this work, we analyse the spectroscopic properties of \wise-selected obscured and unobscured AGN, and use the \OIII\ $\lambda$5007 profile to explore their outflow properties. Though we are still limited by the available spectroscopic datasets, by starting with a large, uniformly selected sample of obscured and unobscured AGN and controlling for known systematic biases, we can shed additional light on the obscured population.  

We adopt a cosmology of $H_0 = 71$ km/s/Mpc, $\Omega_m = 0.27$, and $\Omega_{\Lambda} = 0.73$ throughout.

\section{DATA} \label{sec:data}
\subsection{\wise-selected AGN} \label{sec:wise_agn}
We assemble obscured and unobscured AGN samples following \citet{2017MNRAS.469.4630D}, and refer the reader there for complete details.  Briefly, we start by selecting AGN candidates from the \wise\ survey catalog data, utilizing a simple color-cut of $W1 - W2 > 0.8$ and magnitude cut of $W2 < 15.05$ \citep[Vega, or 0.16 mJy;][]{2012ApJ...753...30S}, where $W1$ and $W2$ represent the 3.6 and 4.5 $\mu$m bands, respectively.  Following \citet{2016MNRAS.456..924D} and \citet{2017MNRAS.469.4630D}, we only include sources satisfying these criteria in both the ``All-sky''\footnote{\url{http://wise2.ipac.caltech.edu/docs/release/allsky/expsup/}} and ``ALLWISE''\footnote{\url{http://wise2.ipac.caltech.edu/docs/release/allwise/expsup/}} catalogs, in order to provide the cleanest and most conservative sample.  Regions of potentially problematic \wise\ data are excluded, as described in detail in \citet{2017MNRAS.469.4630D}.  These include regions of high Galactic extinction and Moon contamination, flagged \wise\ data (such as optical ghosts and other artifacts), and poor photometric quality.

These sources are then limited to the Sloan Digital Sky Survey \citep[SDSS;][]{2000AJ....120.1579Y} photometric footprint, as complementary optical data are needed to split IR-selected AGN into obscured and unobscured populations.  We match to the SDSS eighth data release\footnote{We continue utilizing the catalogs of \citet{2014MNRAS.442.3443D, 2016MNRAS.456..924D}, used in a series of papers on \wise-selected AGN, and note that no significant changes in this catalog result in matching to a more recent SDSS photometry release.} \citep[DR8;][]{2011ApJS..193...29A} photometry with a radius of 2''.  Since we will be utilizing optical spectroscopic measurements (below), sources without optical counterparts are discarded. Removing optical non-detections does not have a strong impact on other obscured AGN measurements \citep[e.g.][]{2017MNRAS.469.4630D}, though we cannot be certain it does not impact this work.  At the redshifts of our final sample ($<$0.4, see below), \citet[][]{2017ApJ...849...53H} identify $u - W3$ colors as optimal to distinguish between obscured and unobscured AGN, with a $W2-W3$ dependent split of $(u-W3\ \textrm{[AB]} > 1.4(W1-W2\ \textrm{[Vega]}) + 3.2$ used to define obscured sources. We utilize SDSS-provided Galactic extinction corrections and the transformations provided with the \wise\ catalogs and by \citet{2007AJ....133..734B} to convert between Vega and AB magnitude systems. We will discuss the impact of other optical-IR color-cuts (e.g.\ $r-W2$) on our results in Section~\ref{sec:disc}. A very small number of sources with spectroscopic measurements (below) are undetected in the $u$-band, and these are considered obscured.

\subsection{Optical spectra}
To probe outflows in the NLR of \wise\ AGN, we utilize the publicly available AGN Line Profile And Kinematics Archive \citep[ALPAKA\footnote{\url{https://sites.google.com/site/sdssalpaka/home}};][]{2013MNRAS.433..622M}. This catalog provides fits to optical SDSS spectra covering the \OIII\ $\lambda$4959, $\lambda$5007 doublet for a sample of 24,264 optically selected AGN (including both spectroscopically classified Type 1 and Type 2 sources). We refer the reader to \citet{2013MNRAS.433..622M} for full details of this catalog and the spectral fits, but provide a brief summary here.

The catalog starts with all extragalactic spectroscopic sources in the SDSS DR7 catalog \citep[][]{2009ApJS..182..543A} with \OIII\ $\lambda$5007, [NII] $\lambda$6584, H$\alpha$ and H$\beta$ emission detected at $>3\sigma$ confidence in the automated SDSS fitting routine.  Requiring these emission lines limits the sample to $z<0.4$.  A combination of line widths and emission-line diagnostics are then used to identify potential AGN that are re-fit in a more robust way than the SDSS pipeline.  In brief, the fits use multiple Gaussian (narrow and broad) components to fit the SDSS continuum-subtracted (using a 5th order polyonomial) profiles of the emission lines. In the case of \OIII, both the $\lambda$4959 and $\lambda$5007 lines are fit simultaneously with two Gaussians each, with their flux ratios fixed at 3 and the velocity offsets and widths of their respective broad and narrow components tied.  Sources initially classified as AGN based on the SDSS pipeline fits, but later classified as non-AGN with the improved ALPAKA fits, are included in the final ALPAKA catalog. Finally, the catalog includes Type 1/2 classifications based on these emission line fits --- sources with a significant ($>50$\% of the line flux) broad ($>600$ km/s) component in the H$\alpha$ line are considered Type 1, and sources without broad lines but satisfying AGN emission-line diagnostics \citep[e.g.][]{2003MNRAS.346.1055K} are considered Type 2.

We find 5,314 matches between our \wise\ AGN sample and the ALPAKA catalog (1,392 obscured and 3,922 unobscured, or 1.7\% and 3.1\% of the \wise-selected samples with an optical counterpart in the spectroscopic footprint, respectively). The common area for the two samples is $\sim$7,500 deg$^2$.  A small subset of these matches ($\sim$3\% in each subset) do not have reliable fits to the \OIII\ lines or enough reliable photometry to perform SED fits to determine independent AGN luminosities (Section~\ref{sec:lum}), and these are discarded leaving 1,366 obscured and 3,772 unobscured objects. This obscured/unobscured ratio of 27\%/73\% in the sources with optical spectra is similar to the overall ratio in the full \wise-selected population once objects without optical counterparts are removed from the obscured sample.

\subsection{Redshift matching} \label{sec:z_match}
\begin{figure}[!h]
   \centering
   \includegraphics[width=8.5cm]{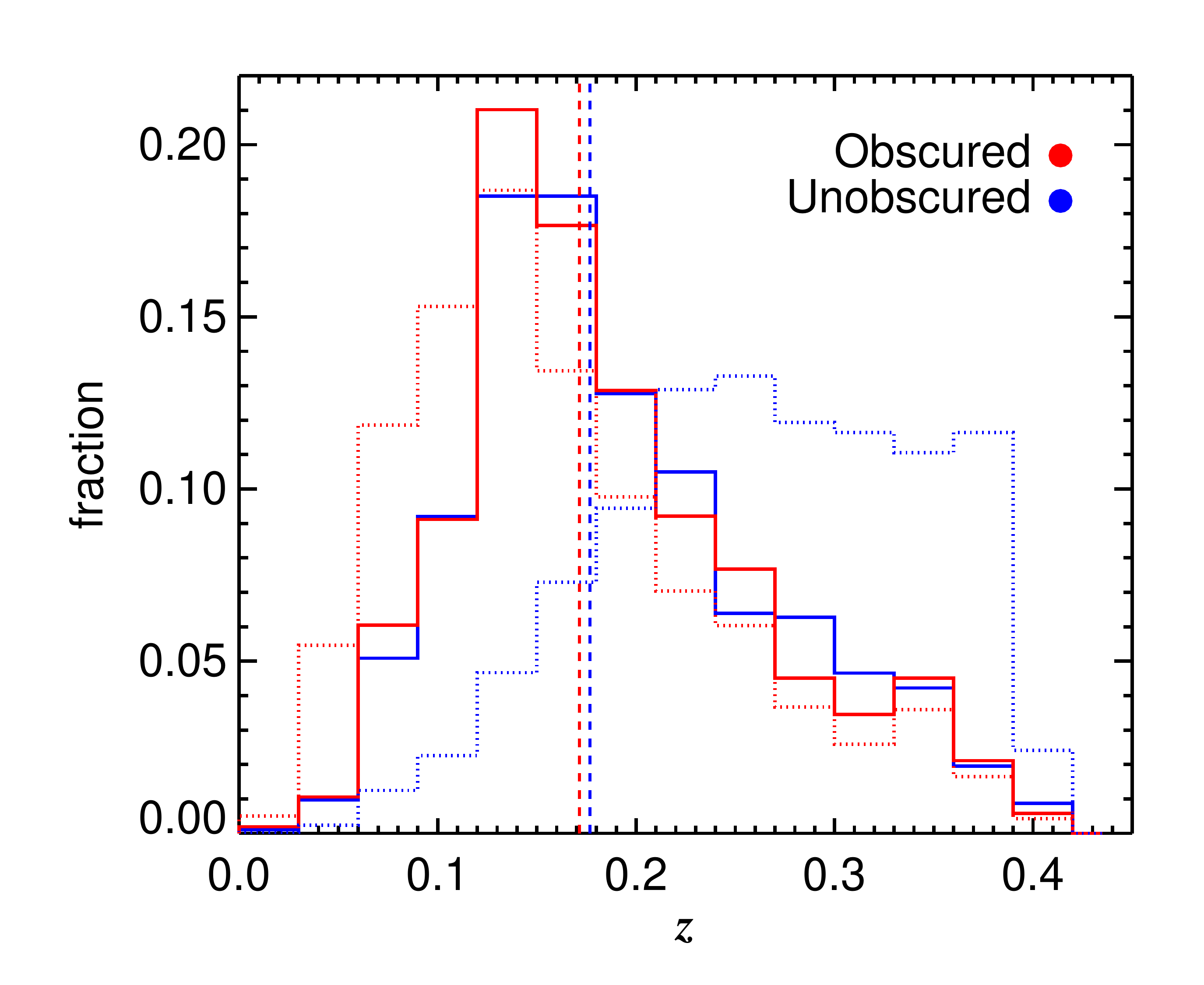}
  \caption{The redshift distributions of the obscured (red) and unobscured (blue) \wise-selected AGN with spectra in the ALPAKA catalog.  The dotted lines show the redshift distributions of all sources with spectra in the ALPAKA catalog, and the solid lines show the distributions of our representative sub-samples selected to match in redshift (see Section~\ref{sec:z_match}). Vertical dashed lines mark the median redshift of each sample, both at $\sim$0.17.\label{fig:z}}
\end{figure}

The dotted red and blue lines in Figure~\ref{fig:z} show the redshift distributions of the \wise-ALPAKA matched obscured and unobscured samples, respectively.  While the full populations of \wise-selected unobscured and obscured AGN are generally well-matched in redshift \citep[based on deep spectroscopic fields and angular cross-correlations;][]{2011ApJ...731..117H, 2017MNRAS.469.4630D}, the additional effects imposed by the SDSS spectroscopic selection lead to a redshift mis-match with obscured sources at preferentially lower redshift.  Before comparing the properties of the two samples we randomly sample them to have matched redshift distributions.

In absolute numbers (rather than the fractions shown in the figure), there are more obscured than unobscured sources up to $z \sim 0.14$.  Below this value, we randomly downsample the obscured sample to match the unobscured distribution, and retain all of the unobscured sources.  At redshifts greater than 0.14, we downsample the unobscured sources to match the obscured distribution, and keep all obscured sources.  To account for the additional variance this downsampling produces, we repeat this process 100 times to provide many redshift-matched samples and average over these when calculating population-level properties (e.g.\ median \OIII\ FWHM), and include the standard deviation among these 100 samples in our error analysis.  We also select a single representative obscured and unobscured sample with median \OIII\ properties consistent with the median of all random subsamples in order to assemble composite spectra (Section~\ref{sec:comps}), visualize distributions (e.g.\ Section~\ref{sec:lum}), and perform additional error analysis (Section~\ref{sec:dists}). These include 1,042 obscured and 924 unobscured AGN, and their redshift distributions are shown with solid lines in Figure~\ref{fig:z}. The median redshift of each is $\sim$0.17.

\subsection{Spectroscopic vs.\ photometric classification} \label{sec:compare}

Given the two methods of classification available for this sample, a comparison of differences between photometric (obscured and unobscured, split at $(u-W3\ \textrm{[AB]} = 1.4(W1-W2\ \textrm{[Vega]}) + 3.2$) and spectroscopic (Type 1 and Type 2, defined by their optical emission line properties) classifications is in order as there is overlap between them. In the redshift-matched obscured samples, the average fraction of spectroscopic Type 2, spectroscopic Type 1, and non-AGN is 57\%, 39\%, 4\%, respectively. Similarly, for the unobscured AGN subsamples, average fractions are 4\%, 93\%, and 3\%, respectively.  We include objects not spectroscopically classified as AGN in the ALPAKA catalog (i.e.\ they satisfied AGN line-width or emission-line diagnostic tests with SDSS pipeline measurements, but not after more detailed line fitting) in this analysis since we are interested in the properties of the \wise-selected populations, but excluding them does not impact the results. 

\begin{figure}[!h]
   \centering
   \includegraphics[width=8.7cm]{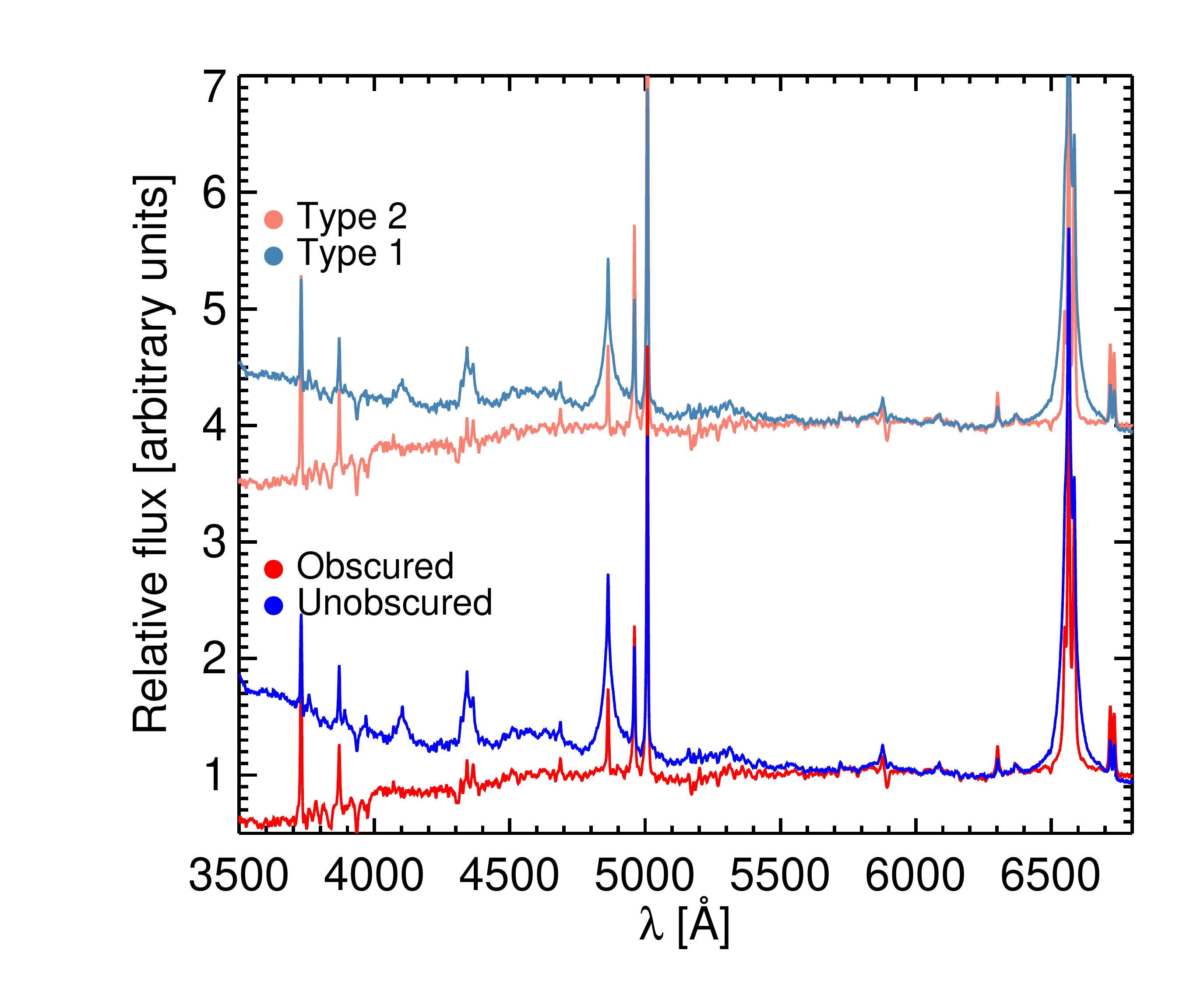}
  \caption{Median composite rest-frame spectra of \wise-selected obscured and unobscured AGN compared to spectroscopic Type 1 and 2 sources (all sources with $z < 0.3$, to show complete coverage of the H$\alpha$ region), normalized at 6100-6300 \AA. The spectroscopic samples are offset vertically for clarity. The spectra are similar, but the mixing of spectroscopic types in the photometrically defined samples is visible, with the unobscured broad H$\alpha$ and H$\beta$ emission lines clearly narrower than the Type 1 lines, and vice-versa for the obscured/Type 2 sources. The differences in the continua in the two classifications are similar.\label{fig:comp_type}}
\vspace{0.2cm}
\end{figure}

To aid in this comparison, in Figure~\ref{fig:comp_type} we show composite spectra (median combined and normalized between 6100 and 6300 \AA\, rest-frame) of all the spectroscopic Type 1/2 sources in the ALPAKA catalog and the \wise-selected obscured/unobscured sources in this analysis (both samples are limited to $z<0.3$ to include the entire H$\alpha$ region in all sources).  It is clear that the unobscured sources have slightly narrower Balmer lines compared to Type 1 sources, and vice-versa for the obscured/Type 2 sources, highlighting the mixing introduced in the photometric classification. However, H$\beta$ is clearly dominated by the narrow component in the obscured sources, while H$\alpha$ retains a significant broad component, though it is narrower than in the unobscured sample. This indicates that there is a significant fraction of the obscured population with only moderate column densities along the line of sight to the broad line region. Because the ALPAKA catalog used H$\alpha$ to determine AGN types, these ``intermediate'' or partially obscured sources (i.e.\ Type 1.8 or 1.9) are grouped into the Type 1 sample while they are more likely to be photometrically classified as obscured. In both classifications, the continua of the unobscured/Type 1 sources is clearly quite blue and ``AGN-like'', while the obscured/Type 2 sources are dominated by galaxy starlight. Based on this analysis (see also Section~\ref{sec:red}), it is likely that our obscured sample includes a significant number of reddened Type 1 AGN, a population of considerable interest in the literature \citep{Richards:2003p3750, 2009ApJ...698.1095U, 2007ApJ...667..673G, 2013ApJ...778..127G, 2015MNRAS.453.3932R}. 

While using optical-IR color to identify obscured AGN at these redshifts includes a significant fraction of broad-line AGN (spectroscopic Type 1 sources), it is clear that objects classified in this way have at least moderate amounts of nuclear obscuration, based on the shape of the continuum and width of the H$\beta$ line in the composite spectra (Figure~\ref{fig:comp_type}).

\section{BOLOMETRIC LUMINOSITIES} \label{sec:lum}
\begin{figure}[!h]
   \centering
   \includegraphics[width=8.cm]{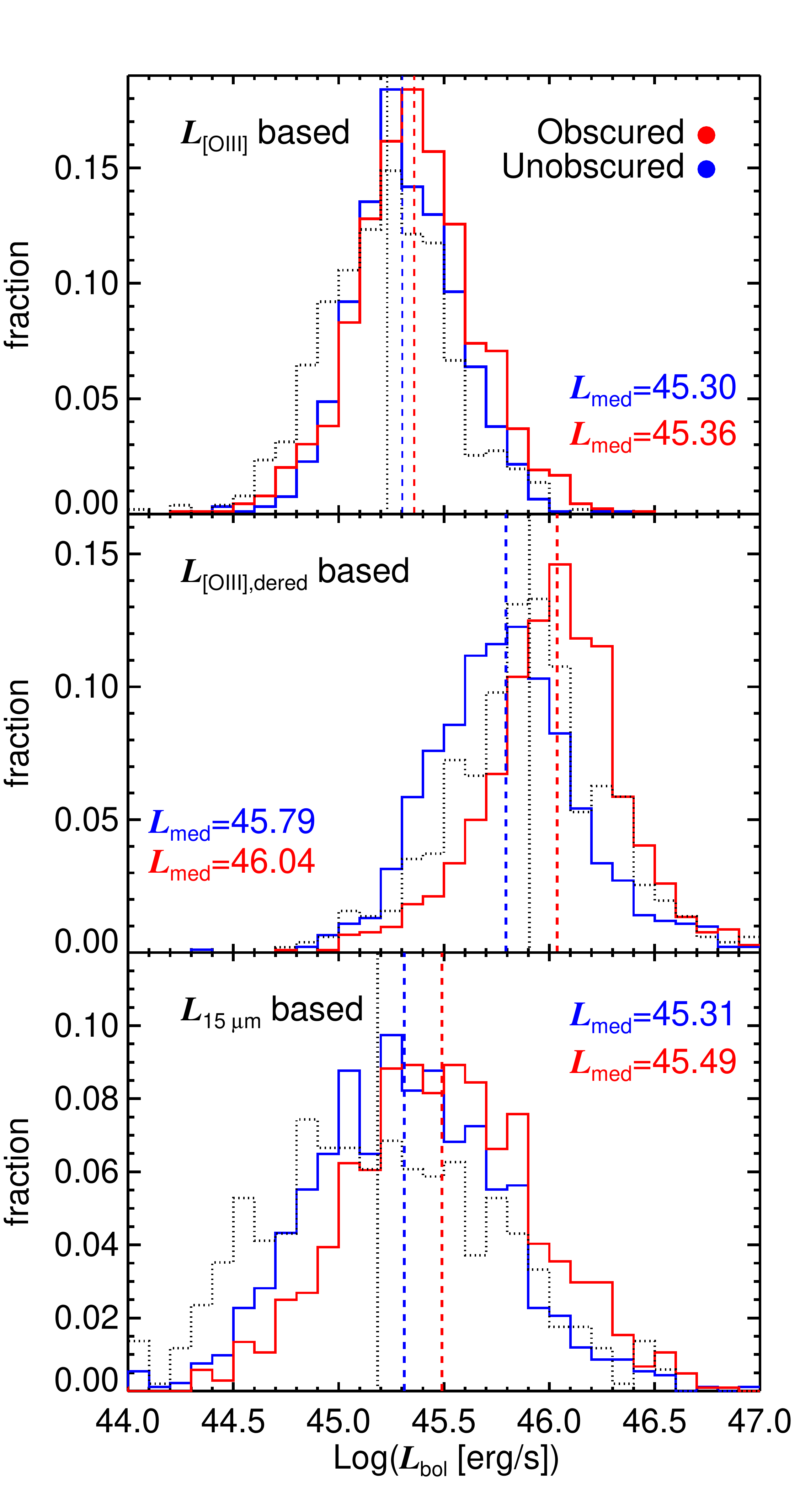}
  \caption{Bolometric luminosities based on measured $L_{\textrm{[OIII]}}$ (top) and dereddened $L_{\textrm{[OIII]}}$ (middle), using the BC of \citet{2017MNRAS.468.1433P}. Red and blue lines show obscured and unobscured \wise\ AGN, while the dotted black line shows the full ALPAKA AGN sample for comparison. The bottom panel shows the bolometric luminosities based on 15 micron AGN luminosities from optical-IR SED fits, using a BC of 14, chosen such that the unobscured median \lbol\ is approximately equal to that of the uncorrected unobscured $L_{\textrm{[OIII]}}$-based \lbol\ (top panel). Vertical lines mark the sample medians, and these are listed in each panel.\label{fig:lbol}}
\vspace{0.2cm}
\end{figure}

While we now have a sample of unobscured and obscured AGN with measured spectral properties matched in redshift, another key parameter to consider is the bolometric luminosity 
\lbol, because the profiles of many emission lines, including \OIII, depend on luminosity \citep[e.g.][]{1977ApJ...214..679B, 2011ApJ...737...71Z}.  The \OIII\ $\lambda$5007 line is often used as a proxy for \lbol\ due to its strength, relative lack of contamination from star formation, and isotropic nature. Many bolometric corrections (BC) for \OIII\ emission-line luminosities have been developed \citep[e.g.][]{2014ARA&A..52..589H}.  Here, we adopt the recent \OIII\ BC of \citet[][the errors on these parameters are excluded for brevity, but the RMS uncertainty on the correction is $\sim$0.4 dex]{2017MNRAS.468.1433P}:
\begin{equation}
\log L_{\textrm{bol}} = 0.56 \log L_{\textrm{[OIII]}} + 22.19.
\end{equation}
Utilizing the luminosity of \OIII\ $\lambda$5007 from the ALPAKA catalog, we show the bolometric luminosities of our samples in the top panel of Figure~\ref{fig:lbol}, along with the full ALPAKA AGN sample for comparison.  In this case, it appears that the bolometric luminosities of the \wise\ samples are well-matched, both at a median \lbol\ $\sim 10^{45.3}$ erg/s.  However, large-scale (kpc) dust obscuration in the host galaxy \citep[e.g.][]{2012ApJ...755....5G} can reduce the flux from the NLR, leading to underestimates of the \OIII\ luminosity. The ALPAKA catalog includes dereddened \OIII\ luminosities, corrected using the Balmer decrement of the narrow components of H$\alpha$ and H$\beta$, and bolometric luminosities based on these are shown in the center panel of Figure~\ref{fig:lbol}.  Here, the median luminosities of the samples are increased significantly, and the obscured sample appears more luminous than the unobscured by $\sim$0.2 dex.  

This shift suggests more dust reddening in the host galaxies of the obscured population. However, it is important to note the difficulty in estimating accurate Balmer decrements in survey-quality optical spectra, compounded by the additional need to decompose the lines into narrow and other components, particularly in the unobscured sample. \citet{2017MNRAS.468.1433P} opted to exclude such a reddening correction to their \OIII\ analysis, even with significantly higher quality spectra, due to the complications of decomposing the lines.  Additionally, the increased median bolometric luminosity in the unobscured sample of $\sim$0.5 dex corresponds to an E($B-V$) of 0.35, or $A_V \approx 1$, quite a large amount for sources that generally still have strong, blue AGN continua. This suggests that the extinction corrections based on the Balmer decrement may be unreliable in these samples.

To avoid the complications of host galaxy extinction in our \lbol\ estimates, as well as provide a measure that is independent of the \OIII\ emission line (the kinematics of which we are studying), we turn to the infrared.  We fit a superposition of four templates \citep[AGN, Sbc/star-forming galaxy, elliptical/passive galaxy, and irregular/starburst galaxy;][]{2010ApJ...713..970A} to the SDSS optical ($u,g,r,i,z$), near-IR \citep[$Y, J, H ,K$ where available from the UKIDSS LAS, $\sim$30\% of the sample;][]{2007MNRAS.379.1599L}, and wise mid-IR ($W1, W2, W3, W4$) SEDs (Carroll et al.\ in prep).  From these, we determine the rest-frame 15 $\mu$m luminosity contributed by the AGN component.  We choose 15 $\mu$m as this roughly corresponds to the longest wavelength \wise\ $W4$ band at 22 $\mu$m at the maximum redshift of our sample, noting that all of our sources (using the representative samples after matching in redshift) are detected in all four \wise\ bands. 

The BC at 15 $\mu$m is $\sim$10 \citep{Richards:2006p3932}. Using this BC provides \lbol\ values that are lower, on average, than those of even the uncorrected \OIII\ estimates (by $\sim$0.2 dex), potentially due to differences in the sample selection and properties compared to those of \citet[][and other samples used to derive BCs]{Richards:2006p3932}.  Therefore, since we are interested in the relative scaling of the luminosities of the obscured and unobscured samples compared to each other, we adopt an IR BC $=14$ \citep[consistent with the range in][]{2012MNRAS.426.2677R} that provides approximately the same median \lbol\ as the uncorrected \OIII\ median luminosity in the unobscured sample. The SED-derived bolometric luminosity distributions are shown in the bottom panel of Figure~\ref{fig:lbol}, and the dashed blue lines in the top and bottom panels show that both approaches produce median unobscured luminosities of $\sim$$10^{45.3}$. Using the SED-based \lbol\ the obscured sample is more luminous by about 0.2 dex compared to the unobscured sample.  It is important to highlight that the luminosities of the full IR-selected obscured and unobscured populations are similar \citep{2007ApJ...671.1365H, 2017MNRAS.469.4630D}, and so this difference is due to selection effects in the SDSS spectroscopic targeting, such as increased host light contamination at lower luminosities in the obscured sample.

Considering these differences in luminosity, we will only make comparisons in matched luminosity bins.  We will also utilize the de-reddened \OIII-derived luminosities to ensure that our results are robust against the details of the luminosity measurement and bolometric corrections.

\section{\OIII\ PROFILES} \label{sec:oiii}

\subsection{Composite Spectra} \label{sec:comps}
In order to gain qualitative insight into the average properties of the \OIII\ emission lines in the obscured and unobscured samples, we first compare composite spectra.  Starting with the continuum-subtracted SDSS spectra, we do a simple median combine with no outlier rejection (including various rejection types makes a negligible impact on the results).  Since the spectra are already continuum subtracted, no normalization is applied prior to combination. This does lead to more scatter in the peaks of the lines due to the range of luminosities.  To generate errors on the composites, we randomly sample (with replacement, such that individual sources can appear more than once in a given subsample) a new set of spectra of the same size as the data set being considered and create a new composite N=1000 times, and take the standard deviation of these composites at each wavelength.

\begin{figure}[!h]
   \centering
   \includegraphics[width=8.5cm]{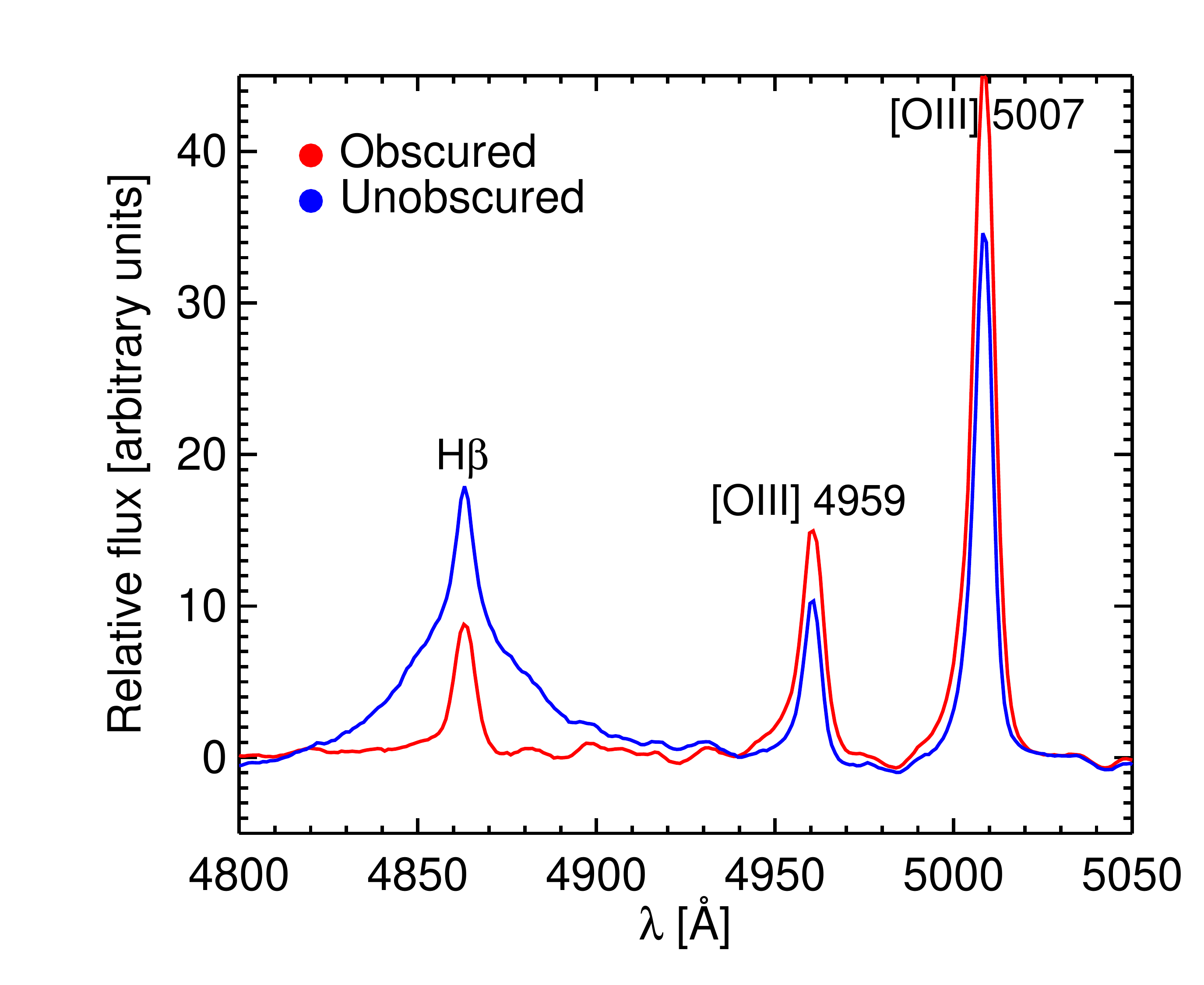}
  \caption{Median composite continuum-subtracted spectra of the obscured and unobscured H$\beta$ and \OIII\ region used in this anlaysis. The unobscured sample clearly has a much stronger broad H$\beta$ component.  The \OIII\ line shows clear blueshifted broad components in both samples, more so in the obscured sample. This is likely explained by the difference in average luminosity for the full samples, and the fact that luminosity correlates with outflow velocity in AGN. Errors on these composites are omitted for clarity.\label{fig:comp_full}}
\vspace{0.2cm}
\end{figure}

First, we show the median spectra of the full H$\beta$/\OIII\ complex in Figure~\ref{fig:comp_full} (errors here are omitted for clarity).  It is clear from the H$\beta$ line that the unobscured sample has a much stronger broad component than the obscured sample, despite the disagreement in photometric and spectroscopic classification for a significant fraction of the objects. There is also a clear broad, blueshifted wing in both of the \OIII\ lines, which is broader in the obscured sample.  In spectroscopic samples, the Type 2 AGN tend to have similar or less prominent broad blue wings \citep{2013MNRAS.433..622M}.  However, we note that the difference seen here is possibly due to the difference in median luminosity of the samples, considering the correlation between the broad component of \OIII\ and luminosity (below).

\begin{figure}[!h]
   \centering
   \includegraphics[width=8.5cm]{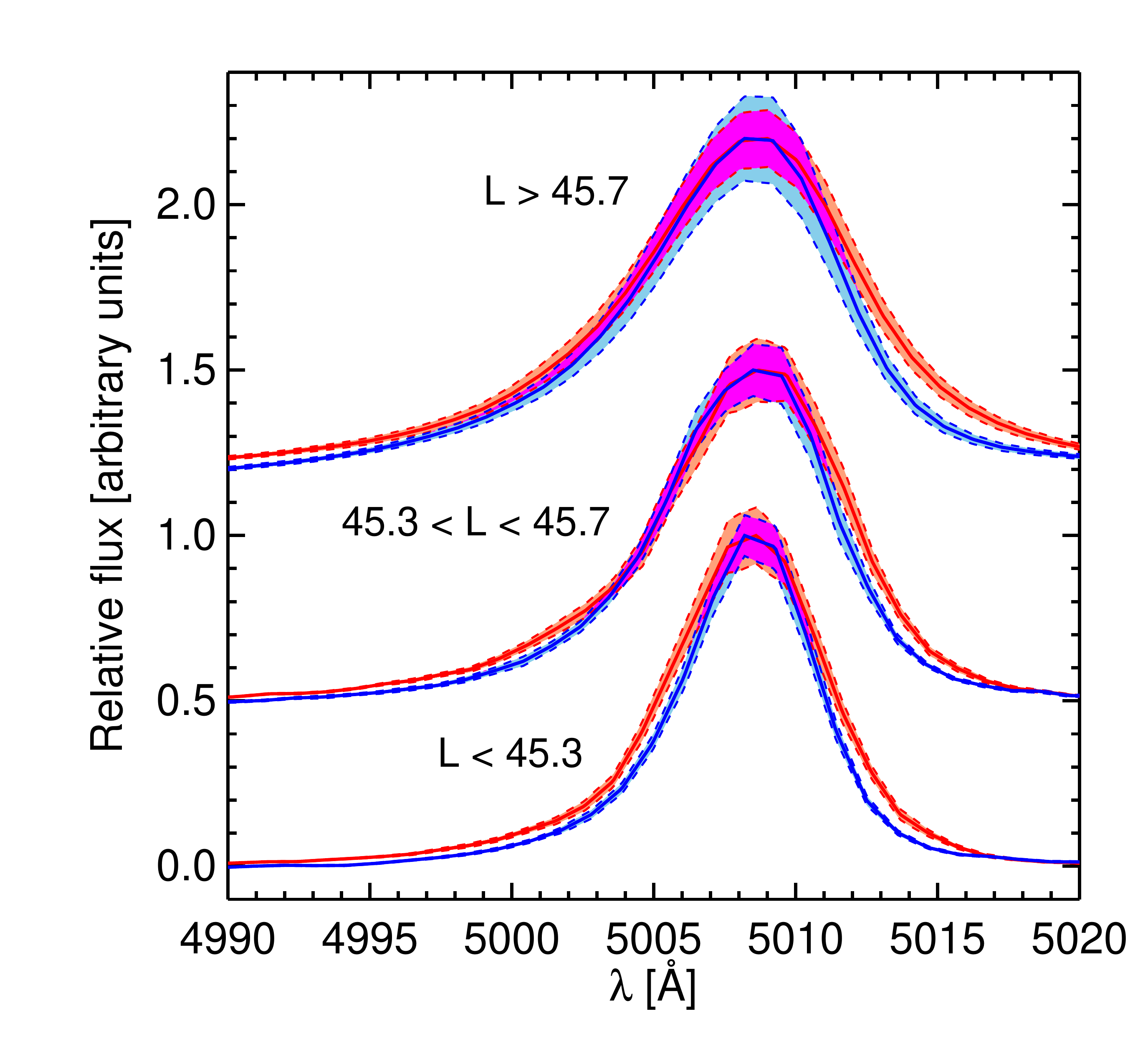}
   
  \caption{Composite spectra of the \OIII\ $\lambda$5007 line binned by luminosity. Error ranges are indicated by dashed lines and filled colors.  In both samples, there is a clear trend toward a stronger and wider broad component with higher luminosity. This component is generally more blueshifted as well, though in the highest luminosity bin  this trend may not hold in the obscured sample (but it does when looking at sample distributions).  In general, the obscured sample has a wider broad component indicative of higher velocity outflows. Table~\ref{tbl:results} contains best-fit parameters to these composites, as well as composites binned using \OIII-based \lbol\ values (not shown here as they are qualitatively similar).\label{fig:comp_lum}}
\vspace{0.2cm}
\end{figure}

In order to account for the difference in average luminosity, we break the samples into three \lbol\ bins, with boundaries defined to split the obscured sample roughly evenly ($\sim$350 objects per bin; see Table~\ref{tbl:results}): $\log$ \lbol\ $< 45.3$, $45.3 < \log$ \lbol\ $< 45.7$, and $\log$ \lbol\ $>45.7$ for bolometric luminosities determined from $L_{15 \mu m}$. For comparison, we also split the sample based on de-reddened $L_{\textrm{[OIII]}}$ luminosity, and since these values tend to be higher the divisions are at $\log$ \lbol\ $< 45.9$, $45.9 < \log$ \lbol\ $< 46.15$, and $\log$ \lbol\ $>46.15$. The composites of just the \OIII\ $\lambda$5007 line in these bins are shown in Figure~\ref{fig:comp_lum} --- the composites using the \OIII-based bolometric luminosities are similar and so are not shown.  It is clear that the broad, blueshifted component of \OIII\ gets stronger and wider with luminosity in both samples, in agreement with previous work (see Section~\ref{sec:intro}) and suggestive of radiatively driven winds in both obscured and unobscured \wise\ AGN.  However, it is also apparent that in all luminosity bins, regardless of how the bolometric luminosity is determined, the obscured sources have wider broad components, potentially more blueshifted (though the lines are qualitatively more symmetric in the obscured sample), indicative of stronger outflows.

To provide some quantitative comparison of the composite spectra, we also perform fits to the \OIII\ $\lambda$4959 and $\lambda$5007 lines in the composites, following a similar methodology as \citet{2013MNRAS.433..622M} using the \textsc{mpfitfun} IDL routine. The results of these fits are listed in Table~\ref{tbl:results} (values from composite fits are listed with a ``c'' subscript). The top half shows results based on SED-derived \lbol, and the bottom half shows the results using dereddened \OIII-based \lbol. These results verify that the obscured sample has both wider (indicated by the FWHM) and more blueshifted (indicated by the velocity offset $v$, which we always list as positive since all are in the same direction) broad components. The exception to this is the highest luminosity bin where the composite obscured \OIII\ profile appears more symmetric (while remaining wider), leading to lower blueshift values. However, in the next section we show that this is not true when considering the distributions of sources in each bin, suggesting that the composite is being affected by outliers in the wings of the line.

\subsection{Median \OIII\ properties} \label{sec:dists}
While the composite spectra are useful for qualitative comparisons, the precise details of creating them can complicate quantitative comparisons.  We therefore focus the remainder of this work on the distributions of \OIII\ parameters in each sample, utilizing the same bins in luminosity as in Section~\ref{sec:comps}.  In cases where no broad component was necessary in fitting the lines \citep{2013MNRAS.433..622M}, we use the properties of the narrow \OIII\ lines as limits on the broad components, though simply neglecting these sources does not change the general results or conclusions.

In calculating the median of any property, we first calculate the median of each of the 100 obscured and unobscured redshift-matched subsamples, and then adopt the median of these as the final value. Errors on median values are generated with a conservative bootstrapping method. We randomly resample each quantity under consideration for each object in accordance with its errors in the individual ALPAKA fits, while simultaneously selecting a random subset of objects (with replacement) of the same size as the initial set N=10,000 times, recalculating the median for each resample. The standard deviation of the resulting distributions of medians is taken as the 1-$\sigma$ error on the median. To account for variance in median properties introduced by resampling to match in redshift, we calculate the standard deviation of the medians of each of the 100 subsamples and add this in quadrature to the bootstrapped errors.

\begin{figure}[!t]
	\centering
   	\vspace{0.1cm}
		\centering
	   	\includegraphics[width=8cm]{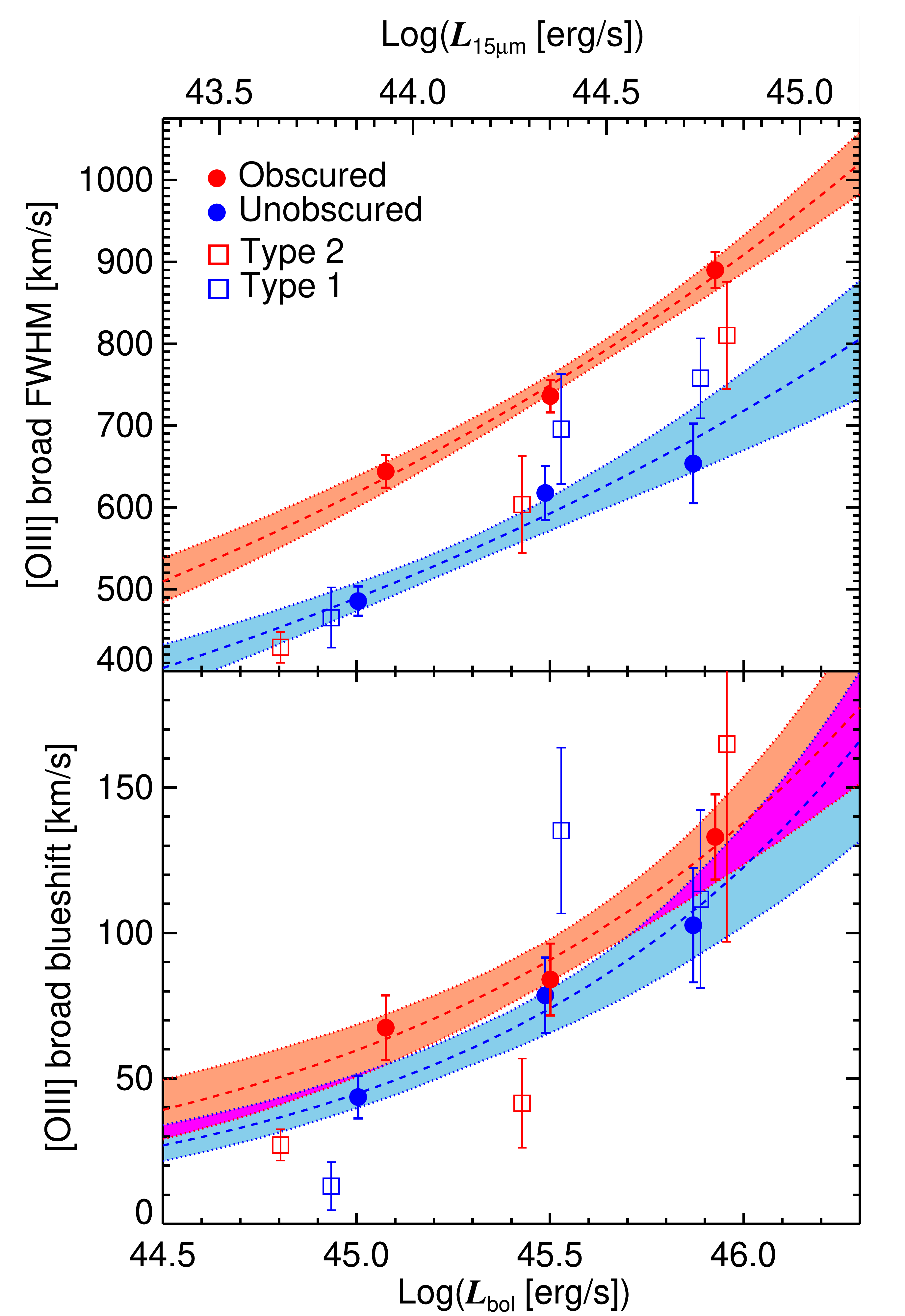}
  	\caption{The broad \OIII\ FWHM (top panel) and blueshift (bottom panel) as a function of bolometric luminosity based on 15 $\mu$m SED-based luminosities. The trends are very similar when using dereddened \OIII-based \lbol\ bins. Bins in luminosity are the same as used for the composite spectra (Figure~\ref{fig:comp_lum}), and listed in Table~\ref{tbl:results}. Both the FWHM and the blueshift increase with luminosity for the obscured and unobscured AGN.  These relationships are well-fit by power laws (Equation~\ref{eq:plaws} and Table~\ref{tbl:fits}), and the obscured source velocities are systematically larger than the unobscured sources. Spectroscopically classified sources from the SDSS (red and blue squares) do not follow this trend.\label{fig:fwhm_l}}%
\vspace{0.2cm}
\end{figure}

In Figure~\ref{fig:fwhm_l} we show both the median broad \OIII\ FWHM (top panel) and broad \OIII\ blueshift ($v$, bottom panel; note that these values are always given as positive, as they are all in the same direction) in each bin of \lbol. Points are placed along the $x$-axes at the median \lbol\ of the sample in each bin, which because of the differences in the overall distributions (Figure~\ref{fig:lbol}) are not always identical.  We also include points for the full ALPAKA AGN spectroscopically-classified Type 1 and 2 samples for comparison.  The median values of the FWHM, $v$, and \lbol\ are all listed in Table~\ref{tbl:results}. The table also includes results using \OIII-based \lbol\ bins, but these are not shown in the figure as they are qualitatively similar. 

It is immediately clear that both samples have increasing velocity widths and blueshifts with luminosity, and the obscured sample has consistently higher velocities in all bins.  We also note that in all bins, the obscured sample has a higher fraction of objects that require a broad component to improve the fit than the unobscured sample, by $\sim$10\%, further evidence of an enhanced broad component in the obscured population.

In order to quantify the differences between the \wise\ obscured and unobscured samples, we parameterize the FWHM - \lbol\ and $v$ - \lbol\ relationships with simple power laws that generally provide good fits to the data:
\begin{equation} \label{eq:plaws}
\textrm{FWHM} = A L_{\textrm{bol}}^{\alpha}
\end{equation}
\begin{equation}
v = B L_{\textrm{bol}}^{\beta}.
\end{equation}
The best fit parameters (using simple $\chi^2$ minimization) are listed in Table~\ref{tbl:fits}.  The error ranges of the fits are produced by randomly resampling each point in accordance with its errors and re-performing the fit 1000 times.  Because the fit parameters are highly covariant, we do not list errors on the individual parameters in the table.  It is clear that the two samples have different normalizations and similar slopes (for the SED-derived luminosities). 

\begin{deluxetable*}{lccccccccccccc}[!h]
\centering
\tabletypesize{\footnotesize}
\tablecaption{\OIII\ properties binned by \lbol \label{tbl:results}}
\tablehead{
 \colhead{} & \multicolumn{6}{c}{Obscured} & \colhead{} & \multicolumn{6}{c}{Unobscured}  \\
	           \cline{2-7}                                    \cline{9-14}                     \\
 \colhead{} &  \colhead{N}  & \colhead{$\langle \log L_{bol} \rangle$} & \colhead{$\langle$FWHM$\rangle$} & \colhead{FWHM$_{\textrm{c}}$}  &  \colhead{$\langle v \rangle$} & \colhead{$v_{\textrm{c}}$} & \colhead{} & 
                       \colhead{N} & \colhead{$\langle \log L_{bol} \rangle$} & \colhead{$\langle$FWHM$\rangle$} & \colhead{FWHM$_{\textrm{c}}$}  &  \colhead{$\langle v \rangle$} & \colhead{$v_{\textrm{c}}$} \\
 \colhead{} &  \colhead{}    &    \colhead{[erg/s]}                                  &          \colhead{[km/s]}                       &             \colhead{[km/s]}                &              \colhead{[km/s]}       &     \colhead{[km/s]}             & \colhead{} &
                      \colhead{}    &    \colhead{[erg/s]}                                  &          \colhead{[km/s]}                       &             \colhead{[km/s]}                &              \colhead{[km/s]}       &     \colhead{[km/s]}             
 }
\startdata
15 $\mu$m-based \lbol    &        &         &       &         &       &      & &         &         &        &        &      &      \\
\cline{1-1} \\
$L_{bol} < 45.3$            & 351 & 45.1 & 643 & 822 & 67 & 128 & & 456 & 45.0 & 486 & 593 & 43 & 78  \\
$45.3 < L_{bol} < 45.7$ & 359 & 45.5 & 735 & 907 & 84 & 138 & & 287 & 45.5 &  617 & 679 & 79 & 119 \\
$L_{bol} > 45.7$            & 332 & 46.9 & 889 & 1058 & 98 & 97 & & 181 & 46.9 & 654 & 773 & 103 & 133 \\ 
\\
\OIII-based \lbol    &        &         &       &         &       &      & &         &         &        &        &      &      \\
\cline{1-1} \\
$L_{bol} < 45.9$            & 333 & 45.7 & 688 & 836 & 75 & 106 & & 586 & 45.6 & 496 & 639 & 44 & 79  \\
$45.9 < L_{bol} < 46.15$ & 354 & 46.0 & 740 & 829 & 87 & 120 & & 191 & 46.0 &  647 & 711 & 114 & 149 \\
$L_{bol} > 46.15$            & 355 & 46.3 & 805 & 998 & 105 & 81 & & 147 & 46.4 & 692 &  787  & 109 & 94
\enddata
\tablecomments{The quantities in brackets represent sample medians, rather than means.  The ``c'' subscript denotes values from fits to the composite spectra, following a similar 
	prescription as \citet{2013MNRAS.433..622M}.  The velocity offsets $v$ are given as positive for simplicity, as they are all blueshifted. Typical errors on the composites are $\sim$50 
	km/s for the FWHM and $\sim$20 km/s for the velocity offset. The median values and their errors are shown in Figure~\ref{fig:fwhm_l}.}
\end{deluxetable*}

\begin{deluxetable*}{lcccccccccc}[!h]
\centering
\tabletypesize{\footnotesize}
\tablecaption{Power law fits to FWHM and $v$ vs. luminosity \label{tbl:fits}}
\tablehead{
            \colhead{}        & & \multicolumn{4}{c}{15 $\mu$m-based \lbol} & \colhead{} & \multicolumn{4}{c}{\OIII-based \lbol} \\
                                               \cline{3-6}                                             \cline{8-11}    \\
	    \colhead{}        & & \colhead{$\log A$} & \colhead{$\alpha$} & \colhead{$\log B$} & \colhead{$\beta$} & \colhead{} & \colhead{$\log A$} & \colhead{$\alpha$} & \colhead{$\log B$} & \colhead{$\beta$}
}
\startdata
Obscured                     & &    -4.74  &  0.17       &   -14.66   &   0.36    &  & -2.47     & 0.12        &   -9.48    & 0.25       \\
Unobscured                  & &    -4.80  & 0.16        &  -18.08   &   0.44    &  & -6.72     & 0.20        &   -21.19  & 0.50    
\enddata
\tablecomments{The fits and their errors for the 15 $\mu$m-based \lbol\ results are shown in Figure~\ref{fig:fwhm_l}.}
\end{deluxetable*}

\subsection{Outflows vs.\ AGN reddening} \label{sec:red}
Given the low redshifts and quality photometry and spectra for this dataset, our SED fitting allows us to explore the \OIII\ outflow properties simply as a function of AGN reddening rather than enforcing a hard split between obscured and unobscured populations.  We note that this is not possible for example for the full \wise\ AGN population, which is why the bulk of the analysis thus far has focused on cuts that are more broadly applicable.  The optical-IR SED modeling includes the $E(B-V)$ of the AGN component, and as expected the reddening in the obscured sample is, on average, larger than the unobscured sample (with medians of 0.74 and 0.03, respectively). While there is no strict definition, if we define a ``red AGN'' as having $0.1 < E(B-V) < 3$ \citep[e.g.][]{2007ApJ...667..673G}, we find that 38\% of our total sample meet this definition, or 41\% of the obscured sample and 37\% of the unobscured sample. This fraction is consistent with the number of broad-lined AGN in the obscured sample.  

\begin{figure}
	\centering
   	\vspace{0.1cm}
		\centering
	   	\includegraphics[width=7.5cm]{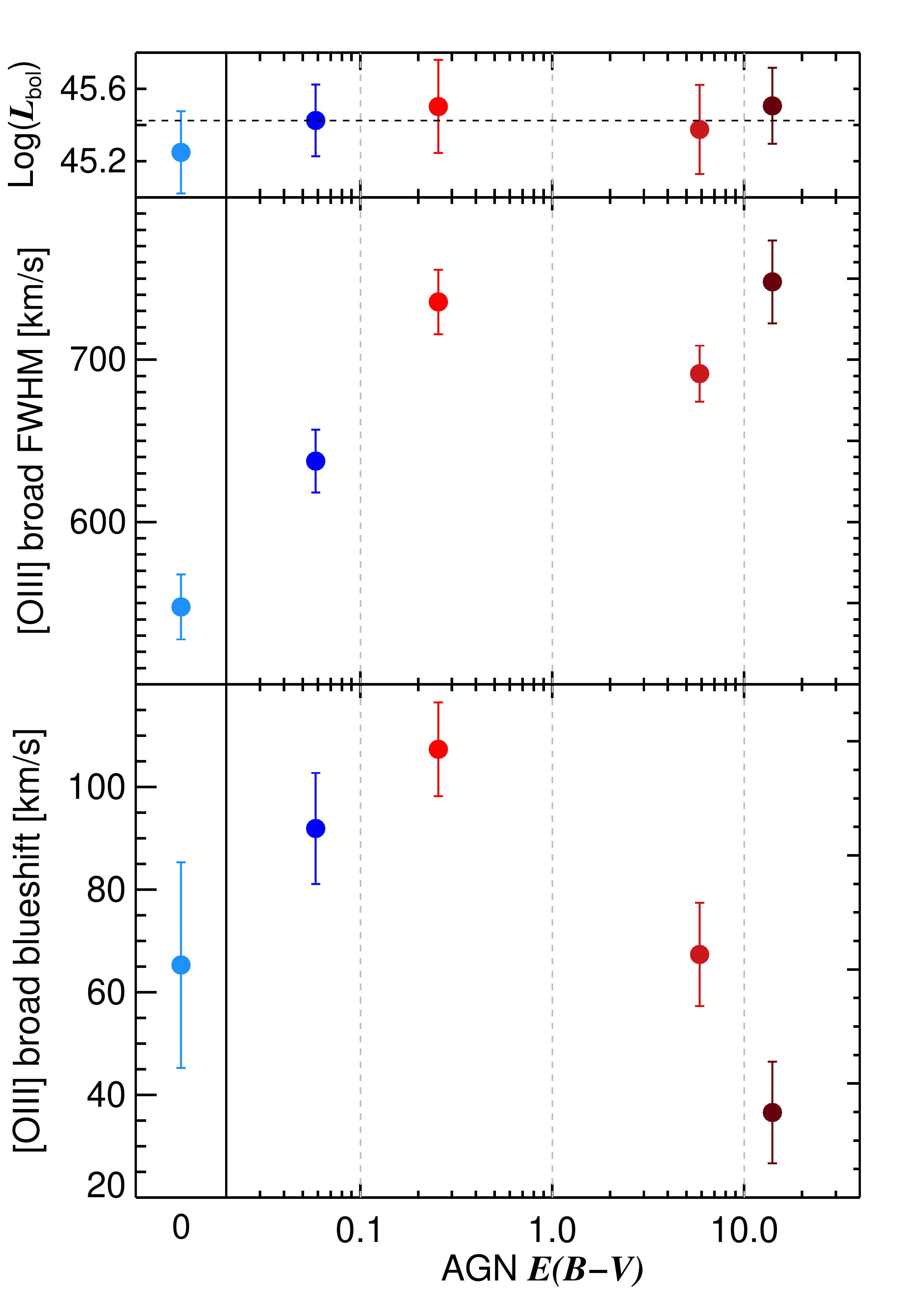}
  	\caption{The broad \OIII\ velocity properties as a function of AGN $E(B-V)$ (from our SED fitting). Bins are evenly spaced in logarithmic $E(B-V)$ space (marked by dashed grey lines), and points are placed at the median value in each bin with colors from light blue to dark red used to indicate obscuration levels. The solid black vertical line separates the panels such that we can include the group with median $E(B-V)=0$ on the logarithmic axis. Median AGN $E(B-V)$ values for our obscured and unobscured samples are 0.74 and 0.03, respectively. The top panel shows the median bolometric luminosity in each bin, with the median of all bins marked with a horizontal dashed line. \label{fig:fwhm_ebv}}%
\vspace{0.2cm}
\end{figure}

In Figure~\ref{fig:fwhm_ebv}, we show the broad \OIII\ FWHM and blueshift in logarithmically even spaced $E(B-V)$ bins. The top panel shows the median bolometric luminosity in each bin, highlighting that there is no significant trend in luminosity with reddening that could impact any trend with FWHM or blueshift.  It is clear that both the broad \OIII\ FWHM and blueshift increase from zero to moderate reddening values, mimicking the difference found earlier between obscured and unobscured sources. However, around $E(B-V) = 0.5$, this trend flattens with FWHM and reverses with blueshift.  We note that the reddening values in the range 0.1 to 1 where this transition occurs are typical of what is found in red quasars at higher redshift, which may be in a specific evolutionary phase \citep[e.g.][]{2008ApJ...674...80U, 2015ApJ...806..218G}. We will discuss the implications of this further in Section~\ref{sec:model}.

\section{DISCUSSION} \label{sec:disc}
Previous studies of the obscured population of AGN have focused on largely optically-selected samples classified based on their spectral properties (Type 1 and 2 AGN).  With large IR surveys like \wise, a new window is open to obscured AGN classified on photometry (obscured and unobscured AGN) --- but these methods of selection and classification lead to distinct population differences. This is important to understand in the context of the next generation of imaging surveys, as well as to fully categorize the roles of orientation and other factors, such as evolution, in AGN obscuration.

Because Type 2 sources are defined as having only narrow emission lines, they must have high column densities toward the broad line region that remove both broad H$\beta$ as well as H$\alpha$.  The photometric selection on the other hand will generally include sources with heavily extincted broad H$\beta$ and red continua, but retain sources at only moderate reddening that still have significant broad H$\alpha$.  There are also luminosity and Eddington ratio differences, with the \wise-selected samples having higher values of both \citep[Figure~\ref{fig:lbol} and e.g.][]{2013ApJ...772...26A}. As we have shown above and in other work, these result in important physical differences that are not seen in spectroscopic samples --- the \wise-selected population of obscured AGN is unique.

In the simplest orientation-based models of AGN, where obscuration occurs in the nuclear regions and viewing angle determines the type of AGN observed, the large-scale properties of obscured and unobscured AGN should be the same.  Many recent works have found that this is not the case, including differences in halo masses \citep[e.g.][]{2017MNRAS.469.4630D}, neighboring galaxies \citep{2014arXiv1411.6735V}, and star-formation/supernova rates \citep{2015ApJ...802...50C, 2017ApJ...837..110V}.  Our analysis supports these findings. First, we identify higher extinction levels in the obscured AGN hosts, as is also found in spectroscopically classified Type 2 AGN \citep{2006AJ....132.1496Z}.  While the obscured AGN are more luminous than the unobscured sample based on their SED fits, their observed \OIII\ luminosities are very similar (Figure~\ref{fig:lbol}), implying large (galaxy) scale dust blocking the extended NLR in the obscured sources.  This could indicate that either the AGN are obscured \emph{because} of the large scale dust (and significant extinction occurs on large scales), or that obscured AGN are more likely to be found in galaxies in a different, more dusty phase.

We have also shown that there are significant differences in the large-scale outflow properties of obscured and unobscured AGN, with the obscured population having higher velocity outflows (as measured by the broad \OIII\ FWHM and velocity offset) at a given AGN luminosity (Figures~\ref{fig:comp_lum} and~\ref{fig:fwhm_l}). The magnitude of this difference is relatively constant with luminosity, and is similar regardless of how \lbol\ is measured. Interestingly, this is not the case for spectroscopically classified Type 1 and 2 sources, which have similar outflow properties or even the reverse trend. Figure~\ref{fig:fwhm_ebv} shows that this may be because there is a turnover in the relationship between outflow velocity and $E(B-V)$. Since spectroscopic Type 2 sources are required to have large extinction values to completely block the broad line region, they will not have the high velocities seen in sources with more moderate $E(B-V)$ values --- indeed, our SED fitting of all the Type 2 AGN in the ALPAKA catalog indicate they have a median $E(B-V) = 14$, as opposed to the median value of 0.74 in the photometrically classified obscured sample. Finally, a recent study by \citet{2017arXiv171002525T} of \OIII\ profiles in \wise\ and SDSS selected dust-obscured galaxies (DOGs), which represent the highest luminosity end of IR-selected AGN, also find strong ionized outflows.  These have velocity offsets of hundreds to a thousand km/s and \lbol $\sim 10^{47}$ erg/s, consistent with the trend we find in our obscured sample. 


\subsection{The impact of radio jets}
Several studies, including \citet[][whose spectral fits make up the basis of much of this work]{2013MNRAS.433..622M} have found that outflow velocity is correlated with radio power, sometimes more so than bolometric luminosity, suggesting that the radio jets are the primary driver of large-scale outflows. In our sample, there are a significant fraction of sources with radio counterparts, and these are preferentially in the obscured population --- 10\% of the unobscured sources have counterparts at 1.4 GHz in either Faint Images of the Sky at Twenty-Centimeters \citep[FIRST, $\sim$0.6 mJy 5$\sigma$ detection limit;][]{Becker:1995p345} or the NRAO VLA Sky Survey \citep[NVSS, $\sim$2.5 mJy 5$\sigma$ detection limit;][]{Condon:1998p1008}, while 40\% of the obscured sources have a radio counterpart \citep{2013MNRAS.433..622M}. The vast majority of detections in both samples are in FIRST. This discrepancy between the two samples is at least partly due to biases imposed by the SDSS spectroscopic selection algorithms, in particular that the obscured sources are more likely to be selected for spectroscopic follow up based on radio emission rather than e.g.\ optical properties \citep[e.g.][]{2004AJ....128.1002Z, 2017ApJ...849...53H}. We perform two checks on the impacts of radio jets on our results.

The first is to simply remove all radio detected sources from the samples and repeat the analysis, though we note that this can have other effects on the sample that may impact outflow properties, such as changing the black hole mass and Eddington ratio distributions. In doing so, the results are not significantly changed, though the median FWHM and blueshift in each sample and \lbol\ bin is reduced slightly.  For example, the obscured FWHM from lowest to highest luminosity become 614, 695, and 848 km/s, a reduction of $\sim$5\% in each bin.  The change is very similar in the unobscured sample, and the general trend is still preserved.

We also apply our same analysis using bins in radio luminosity instead of \lbol\, where we subsample from each bin such that the median bolometric luminosities in each is within 0.25 dex of the others (all at $\log$ \lbol\ $\sim 45.2$ erg/s).  In this case, we find that the unobscured FWHM-$L_{\textrm{1.4GHz}}$ relationship is roughly flat, while for the obscured sample it increases slightly by $\sim$150 km/s from low to high radio luminosity.  However, the obscured FWHM is generally smaller than or consistent with that of the unobscured sample, rather than consistently larger.  We note that controlling for \lbol\ in each bin does reduce the sample size significantly, particularly for the highest radio luminosity bin, and the errors are increased accordingly. The tests performed here imply that while radio jets may play a role in the outflow properties of these samples, it is not the dominant factor, and radiation pressure is needed to produce the enhanced velocities in the obscured sample.

\subsection{Luminosity, Eddington ratio, and bolometric corrections}
While we have matched our samples in luminosity bins, it is possible that other key parameters tied to either the luminosity indicators or the outflows are not matched, such as black hole mass or the Eddington ratio $\lambda_{\textrm{Edd}}$.  Recently it was shown that the obscured fraction, and hence the nuclear dust covering factor, is a function of $\lambda_{\textrm{Edd}}$ in nearby X-ray AGN, with higher Eddington ratio sources less likely to be obscured \citep{2017arXiv170909651R}.  This could imply that we are overestimating the BC for the obscured population. However, in this case applying a smaller BC to the obscured sample implies shifting the relationship in Figure~\ref{fig:fwhm_l} to the left, with higher velocities at lower luminosity, serving to enhance the difference between obscured and unobscured samples further.  A similar argument could be used in the case of a luminosity-dependent BC, as might be expected in receding torus models \citep[e.g.][]{1991MNRAS.252..586L}. 

In addition, when considering radiation pressure on dusty gas, the boost factor (or ratio of effective cross-section to the Thompson cross section) depends on both the column density and the ionization parameter. Since the ionizing flux changes with $\lambda_{\textrm{Edd}}$, so too does the boost factor \citep[e.g.][]{2008MNRAS.385L..43F}. If the obscured sources are preferentially lower Eddington, this would manifest as a lower boost factor, serving to reduce the outflow velocity (all else being equal).

Finally, it is also possible that we underestimate the intrinsic IR luminosity of the obscured sample relative to the unobscured. This could be due to anisotropic emission \citep[e.g.][]{2011ApJ...736...26H}, viewing angle differences, or even extinction in the IR in the most heavily obscured sources --- though the strong, broad H$\alpha$ emission in the obscured composite implies that only a minority of sources are deeply buried in this sample.  However, \citet{2017MNRAS.469.4630D} found that these effects do not seem prevalent in \wise-selected AGN, and in order to completely explain the results here would require us to underestimate the average obscured luminosity by nearly an order of magnitude, while the effects above would impact the luminosities at most by a factor of a few.

\subsection{Physical interpretations} \label{sec:model}
Because we are working with both averages over populations as well as spatially-averaged spectra, modeling the differences we find between the two samples in a detailed way is extremely difficult \citep[e.g.][]{2014MNRAS.442..784Z}. The optical SDSS spectra are from a fiber-fed spectrograph with a fixed 3 arcsecond fiber size, which means that the spectra are spatially averaged over different physical scales. At the median redshift of 0.17, the spectroscopic fibers reach radii of $\sim$4 kpc. 

There are many models of radiation-pressure driven winds in AGN, and the general components are the AGN luminosity, the wind launch radius, the effective cross section (which increases in dusty gas), and the mass profile of the host galaxy \citep[e.g.\ equation 7 of][]{2017ApJ...834...30F}.  These parameters can be considered broadly in the context of our measurements here. 

Since dust enhances the ability of radiation to accelerate a wind due to increased cross section, an angular dependance of the dust content of the wind could lead to orientation affecting the observed velocities.  If the wind is conical in shape but more dusty along the outer edge than in the center, for example due to dust ablated off the edge of the torus \citep[similar to, e.g. the two-walled bi-cone of][]{2017arXiv171000828N}, we could have a scenario in which the outer edge of the conical outflow is moving at higher velocity. If the obscured sources are then seen from preferentially large viewing angles, nearer to the edge of the cone, this could lead to an enhanced asymmetric velocity component.  However, winds launched from the nuclear regions in AGN can have much higher velocities than what we observe here, for example $>10^4$ km/s in broad absorption line quasars.


One simple scenario is that the wind launching radius is different for the obscured and unobscured samples, as gas launched from closer to the nucleus will begin with a higher velocity.  In general evolutionary scenarios for obscured AGN that involve a ``blowout'' phase in which the AGN is clearing its immediate surroundings, the outflowing gas distribution may be more compact and lead to a higher velocity at a given luminosity. Of course, in such a scenario it is expected that the dust content of this cocoon of material is much higher, which can enhance the radiation pressure cross section but also result in more ``dead weight'' behind the material being accelerated \citep{2008MNRAS.385L..43F}.  

Guided by Figure~\ref{fig:fwhm_ebv}, we propose that the need to balance these effects results in the enhanced obscured velocities being dominated by sources with intermediate column densities, explaining why this velocity difference has not been seen in spectroscopically classified samples. In this scenario, not all obscured AGN are in the blowout phase, and the relative numbers of obscured AGN that are spectroscopically classified as Type 1 and Type 2 would correspond to the relative lifetime of such a phase. Based on these fractions discussed Section~\ref{sec:compare}, the blowout phase would correspond to approximately half of the quasar lifetime, roughly consistent with some models and observations \citep[e.g.][]{2008ApJS..175..356H, 2017MNRAS.469.4630D}. If we repeat our analysis using sources in the obscured sample that are also classified spectroscopically as Type 2, the difference in the relationship between luminosity and outflow velocity compared to the unobscured sample is reduced. Conversely, if we limit the obscured sample to only spectroscopic Type 1 sources, the difference is enhanced. This suggests that at least a subset of the obscured population in this work are linked to red AGN, which may be a distinct population tied to certain evolutionary phases such as mergers \citep[e.g.][]{2008ApJ...674...80U, 2015ApJ...806..218G, 2017MNRAS.464.3431H}, though we note that we are probing different redshift ranges and luminosities.

At low redshifts such as these, other color cuts are less efficient at separating obscured and unobscured AGN \citep{2017ApJ...849...53H}. If we instead use an optical-IR color cut of $r-W2 = 6$ to define our obscured and unobscured sources \citep{2007ApJ...671.1365H}, which is commonly used for the full \wise\ AGN population which peaks at $z \sim 1$, we indeed find more mixing of spectroscopic types within the photometrically defined samples. Roughly 50\% of the obscured sources have detected broad lines, and 30\% of the unobscured sample are Type 2 sources.  This additional mixing of reddened Type 1 sources into the obscured sample does enhance the median outflow velocities, as expected in the scenario described above.

\section{SUMMARY \& CONCLUSIONS}
Starting with uniformly selected samples of obscured and unobscured AGN from \wise\, where the classifications are determined purely by optical-IR colors \citep[e.g.][]{2007ApJ...671.1365H, 2017ApJ...849...53H}, we use a subset with SDSS spectroscopy to explore their \OIII\ emission line kinematics and outflow properties.  Using only photometric information to split the obscured and unobscured populations, we find interesting departures from comparisons based on spectral classifications.  We summarize our main results as follows:

\begin{enumerate}
\item Obscured quasars defined photometrically based on optical-IR colors have a mix of spectroscopic types, with nearly half containing broad emission lines, particularly H$\alpha$. The broad component of H$\beta$ is generally much less pronounced than H$\alpha$, and the continua are typically red and galaxy-like. These would be classified as Type 1.8 or 1.9 spectroscopically.

\item Based on SED fitting in the optical and IR, the \wise-selected obscured AGN with spectroscopic follow-up tend to be more luminous, by $\sim$0.2 dex on average, than their unobscured counterparts.  However, this is not reflected in their extended \OIII\ luminosities, suggesting more kiloparsec-scale dust present in the obscured host galaxies.

\item Using both median properties of the broad component of \OIII\ $\lambda$5007 and composite spectra in matched bins of luminosity, we find that obscured AGN have higher velocity outflows than the unobscured population, based on both FWHM and blueshift.  This suggests a fundamental difference between radiatively-driven outflows in \wise-selected obscured quasars compared to their unobscured counterparts that is not seen in optically-selected, spectroscopically-classified samples. At least a subset of the obscured population may be related to the red AGN at higher redshift that are found preferentially in mergers, and could represent a ``blowout'' phase in the AGN lifetime.

\item We divide the samples in terms of nuclear reddening (derived from SED fitting), rather than splitting into obscured and unobscured based on a hard color cut, and we find that the broad \OIII\ FWHM and blueshift increases from AGN $E(B-V) = 0$ to moderate values of $\sim$0.5, before flattening (FWHM) or declining again (blueshift).  This could imply that along the most dusty lines of sight, whether due to orientation or evolutionary state (or some other factor), have too much ``dead weight'' preventing a higher velocity outflow from forming.

\end{enumerate}

With the current data averaged over many spatial scales and looking only at population-level properties, it is not feasible to fully explain this result in any simple outflow model.  Considering the enhanced large-scale extinction, as well as other recent results regarding the star formation, environmental, and dark matter halo properties of the IR-selected obscured population, it is unlikely that simple geometric unification is sufficient to explain these results. Instead, a combination of factors are at play, involving different distributions of outflowing gas and launching radii, potentially related to the evolutionary state of the host galaxies, as well as orientation and projection effects.

\begin{acknowledgements}
We thank the anonymous referee for comments and suggestions that greatly improved the paper. MAD and RCH were partially supported by NSF grant number 1515364 and NASA ADAP award NNX15AP24G.  RCH also acknowledges support from NSF CAREER grant 1554584.

Funding for the SDSS and SDSS-II has been provided by the Alfred P. Sloan Foundation, the Participating Institutions, the National Science Foundation, the U.S. Department of Energy, the National Aeronautics and Space Administration, the Japanese Monbukagakusho, the Max Planck Society, and the Higher Education Funding Council for England. The SDSS Web Site is \url{http://www.sdss.org/}.

The SDSS is managed by the Astrophysical Research Consortium for the Participating Institutions. The Participating Institutions are the American Museum of Natural History, Astrophysical Institute Potsdam, University of Basel, University of Cambridge, Case Western Reserve University, University of Chicago, Drexel University, Fermilab, the Institute for Advanced Study, the Japan Participation Group, Johns Hopkins University, the Joint Institute for Nuclear Astrophysics, the Kavli Institute for Particle Astrophysics and Cosmology, the Korean Scientist Group, the Chinese Academy of Sciences (LAMOST), Los Alamos National Laboratory, the Max-Planck-Institute for Astronomy (MPIA), the Max-Planck-Institute for Astrophysics (MPA), New Mexico State University, Ohio State University, University of Pittsburgh, University of Portsmouth, Princeton University, the United States Naval Observatory, and the University of Washington.

This publication makes use of data products from the Wide-field Infrared Survey Explorer, which is a joint project of the University of California, Los Angeles, and the Jet Propulsion Laboratory/California Institute of Technology, funded by the National Aeronautics and Space Administration.

This work is based in part on data obtained as part of the UKIRT Infrared Deep Sky Survey.

\end{acknowledgements}


\bibliography{/Users/Mike/Dropbox/full_library}

\end{document}